\documentclass[twocolumn,showpacs,aps,prl,superscriptaddress]{revtex4}

\usepackage{graphicx}
\usepackage{dcolumn}
\usepackage{epsfig}
\usepackage{amsmath}

\newcommand{\BaBarYear}    {07}
\newcommand{\BaBarNumber}  {020}

\newcommand{\SLACPubNumber} {12413}
\newcommand{\LANLNumber} {0703038}

 \newcommand{\BaBarType}      {PUB}  

\input pubboard/babarsym
\RequirePackage{xspace}

\newcommand{\calP}{\ensuremath{{\cal P}}}

\newcommand{\pvec}{{\bf p}}

\newcommand{\calB}{\ensuremath{{\cal B}}}

\newcommand{\timesix}{\ensuremath{\times10^{6}}}

\newcommand{\DE}{\ensuremath{\Delta E}}

\newcommand{\mres}{\ensuremath{m_{\rm res}}}
\newcommand{\xf}{\ensuremath{{\cal F}}}
\newcommand{\hel}{\ensuremath{{\cal H}}}
\newcommand{\thetaT}{\ensuremath{\theta_{\rm T}}}
\newcommand{\costhr}{\ensuremath{\cos\thetaT}}

\newcommand\etal{{\it et al.}}
\newcommand{\half}{\ensuremath{{1\over2}}}

\newcommand{\bma}[1]{\boldmath{$#1$}}
\newcommand{\msp}{\ensuremath{\phantom{-}}}

\newcommand{\bfig}{\begin{figure}[htbpc!]}
\newcommand{\efig}{\end{figure}}
\newcommand\bef{\begin{figure}}
\newcommand\edf{\end{figure}}
\newcommand\dbline{\noalign{\vskip 0.10truecm\hrule}\noalign{\vskip 2pt}\noalign{\hrule\vskip 0.10truecm}}
\providecommand{\tbline}{\noalign{\vskip 0.05truecm\hrule\vskip0.05truecm}}

\newcommand\beq{\begin{equation}}
\newcommand\eeq{\end{equation}}
\newcommand\bear{\begin{array}}
\newcommand\enar{\end{array}}
\newcommand\beqa{\begin{eqnarray}}
\newcommand\eeqa{\end{eqnarray}}
\newcommand\ben{\begin{enumerate}}
\newcommand\een{\end{enumerate}}

\newcommand{\UfourS}{\ensuremath{\Upsilon(4S)}}

\newcommand{\etagg}{\ensuremath{\eta_{\gaga}}}
\newcommand{\etappp}{\ensuremath{\eta_{3\pi}}}
\newcommand{\etatogg}{\ensuremath{\eta\ra\gaga}}
\newcommand{\etatoppp}{\ensuremath{\eta\ra\pi^+\pi^-\pi^0}}

   \newcommand{\rhoz}{\ensuremath{\rho^0}}

\newcommand{\fetaKstz}{\ensuremath{\eta K^{*0}}}

\newcommand{\etaKstz}{\ensuremath{\Bz\ra\fetaKstz}}

\newcommand{\fetarhoz}{\ensuremath{\eta\rho^0}}
\newcommand{\etarhoz}{\ensuremath{\Bz\ra\fetarhoz}}
\newcommand{\Betarhoz}{\ensuremath{\calB(\etarhoz)}}
\newcommand{\retarhoz}{\ensuremath{xx^{+xx}_{-xx}\pm xx}}

\newcommand{\uletarhoz}{\ensuremath{xx}}
\newcommand{\ULetarhoz}{\ensuremath{\uletarhoz\times 10^{-6}}}

\newcommand{\setarhoz}{\ensuremath{xx}}
  \newcommand{\fetaggrhoz}{\ensuremath{\eta_{\gamma\gamma} \rho^0}}
  
  \newcommand{\fetappprhoz}{\ensuremath{\eta_{3\pi} \rho^0}}
  
\newcommand{\fetaggfz}{\ensuremath{\eta_{\gamma\gamma} f_0}}
\newcommand{\etafz}{\ensuremath{\Bz\ra\fetafz}}
\newcommand{\fetapppfz}{\ensuremath{\eta_{3\pi} f_0}}
\newcommand{\fetafz}{\ensuremath{\eta f_0}}

\newcommand{\Betafz}{\ensuremath{\calB(\etafz(980))\times\calB(\fzero(980)\to\pipi)}}

\newcommand{\ULetafz}{\ensuremath{\uletafz\times 10^{-6}}}

\newcommand{\azero}{\ensuremath{a_0}}
\newcommand{\azerop}{\ensuremath{a_0(1450)}}
\newcommand{\fzero}{\ensuremath{f_0}}

\newcommand{\Bazeropmpip}{\ensuremath{\calB(\azeropmpip)\times\calB(\azerop^-\to\eta\pim)}}
\newcommand{\BazeropmKp}{\ensuremath{\calB(\azeropmKp)\times\calB(\azerop^-\to\eta\pim)}}

\newcommand{\fazeromKp}{\ensuremath{a_0^- K^+}}

\newcommand{\fazerompip}{\ensuremath{a_0^-\pi^+}}

\newcommand{\fazeropmpip}{\ensuremath{\azerop^-\pip}}
\newcommand{\fazeropmpipgg}{\ensuremath{\azerop_{\gamma\gamma}^-\pip}}
\newcommand{\fazeropmpipppp}{\ensuremath{\azerop_{3\pi}^-\pip}}
\newcommand{\fazeropmKp}{\ensuremath{\azerop^- K^+}}
\newcommand{\fazeropmKpgg}{\ensuremath{\azerop_{\gamma\gamma}^- K^+}}
\newcommand{\fazeropmKpppp}{\ensuremath{\azerop_{3\pi}^- K^+}}
\newcommand{\azeropmpip}{\ensuremath{\Bz\ra\fazeropmpip}}
\newcommand{\azeropmKp}{\ensuremath{\Bz\ra\fazeropmKp}}

\newcommand{\azeromKp}{\ensuremath{\Bz\ra\fazeromKp}}

\newcommand{\azerompip}{\ensuremath{\Bz\ra\fazerompip}}

\newcommand{\BazeromKp}{\ensuremath{\calB(\azeromKp)\times\calB(\azero^-\to\eta\pim)}}
\newcommand{\razeromKp}{\ensuremath{xx^{+xx}_{-xx}\pm xx}}

\newcommand{\ulazeromKp}{\ensuremath{xx}\xspace}
\newcommand{\ULazeromKp}{\ensuremath{\ulazeromKp\times 10^{-6}}\xspace}
\newcommand{\sazeromKp}{\ensuremath{xx}\xspace}
\newcommand{\razeropmKp}{\ensuremath{xx^{+xx}_{-xx}\pm xx}}
\newcommand{\ulazeropmKp}{\ensuremath{xx}\xspace}
\newcommand{\ULazeropmKp}{\ensuremath{\ulazeropmKp\times 10^{-6}}\xspace}
\newcommand{\sazeropmKp}{\ensuremath{xx}\xspace}

\newcommand{\Bazerompip}{\ensuremath{\calB(\azerompip)\times\calB(\azero^-\to\eta\pim)}}
\newcommand{\razerompip}{\ensuremath{xx^{+xx}_{-xx}\pm xx}}

\newcommand{\ulazerompip}{\ensuremath{xx}\xspace}
\newcommand{\ULazerompip}{\ensuremath{\ulazerompip\times 10^{-6}}\xspace}
\newcommand{\sazerompip}{\ensuremath{xx}\xspace}
\newcommand{\razeropmpip}{\ensuremath{xx^{+xx}_{-xx}\pm xx}}
\newcommand{\ulazeropmpip}{\ensuremath{xx}\xspace}
\newcommand{\ULazeropmpip}{\ensuremath{\ulazeropmpip\times 10^{-6}}\xspace}

\providecommand{\msp}{\phantom{$-$}}
\def\Kstar   {\ensuremath{K^*}}

\renewcommand{\razerompip}{\ensuremath{2.1\pm0.8\pm0.3}}
\renewcommand{\ulazerompip}{\ensuremath{3.1}\xspace}
\renewcommand{\sazerompip}{\ensuremath{2.4}\xspace}

\renewcommand{\razeromKp}{\ensuremath{1.0\pm0.6\pm 0.2}}
\renewcommand{\ulazeromKp}{\ensuremath{1.9}\xspace}
\renewcommand{\sazeromKp}{\ensuremath{2.0}\xspace}

\renewcommand{\razeropmpip}{\ensuremath{-3.5\pm2.2\pm0.8}}
\renewcommand{\ulazeropmpip}{\ensuremath{2.3}\xspace}

\renewcommand{\razeropmKp}{\ensuremath{0.8\pm1.5\pm 0.3}}
\renewcommand{\ulazeropmKp}{\ensuremath{3.1}\xspace}
\renewcommand{\sazeropmKp}{\ensuremath{0.6}\xspace}

\renewcommand{\retarhoz}{\ensuremath{0.4\pm0.7\pm 0.2}}
\renewcommand{\uletarhoz}{\ensuremath{1.5}\xspace}
\renewcommand{\setarhoz}{\ensuremath{0.6}\xspace}

\newcommand{\retafz}{\ensuremath{-0.3\pm0.3\pm 0.1}}
\newcommand{\uletafz}{\ensuremath{0.4}\xspace}

\begin{document}

\preprint{\babar-PUB-\BaBarYear/\BaBarNumber} 
\preprint{SLAC-PUB-\SLACPubNumber} 


\begin{flushleft}
\babar-\BaBarType-\BaBarYear/\BaBarNumber \\
SLAC-PUB-\SLACPubNumber\\
hep-ex/\LANLNumber
\end{flushleft}

\par\vskip .2cm

\title{
 \large \bf\boldmath Search for Neutral $B$-Meson Decays to $\azero\pi$,
$\azero K$, $\eta\rhoz$, and $\eta f_0$ 
}

%
\author{B.~Aubert}
\author{M.~Bona}
\author{D.~Boutigny}
\author{Y.~Karyotakis}
\author{J.~P.~Lees}
\author{V.~Poireau}
\author{X.~Prudent}
\author{V.~Tisserand}
\author{A.~Zghiche}
\affiliation{Laboratoire de Physique des Particules, IN2P3/CNRS et Universit\'e de Savoie, F-74941 Annecy-Le-Vieux, France }
\author{J.~Garra~Tico}
\author{E.~Grauges}
\affiliation{Universitat de Barcelona, Facultat de Fisica, Departament ECM, E-08028 Barcelona, Spain }
% ...
\collaboration{The \babar\ Collaboration}
\noaffiliation

\date{\today}

\begin{abstract}
We present a search for \Bz\ decays to charmless final states involving an 
$\eta$ meson, a charged pion and a second charged pion or kaon.
The data sample corresponds to 383\timesix\ \BB\ pairs collected with the
\babar\ detector operating at the PEP-II asymmetric-energy $B$ Factory
at SLAC.  We find no significant signals and 
determine the following 90\% C.L. upper limits:
$\Bazerompip < \ULazerompip$, $\BazeromKp < \ULazeromKp$, 
$\Bazeropmpip < \ULazeropmpip$, $\BazeropmKp < \ULazeropmKp$, 
$\Betarhoz < \ULetarhoz$, and $\Betafz < \ULetafz$. 

\end{abstract}

\pacs{13.25.Hw, 12.15.Hh, 11.30.Er}

\maketitle


We investigate the decays of a \Bz\ meson to final states with an $\eta$ meson, 
a charged pion, and either a second charged pion or kaon.  The 
Dalitz plots for the decays without an intermediate charmed meson are expected 
to have contributions from quasi-two-body decays such as \azerompip, \azeromKp,
\etarhoz, \etafz, and \etaKstz\ \cite{CC}.  The last of these has been 
investigated recently \cite{etakstpub} so in this paper we concentrate on the
others, where \azero\ is either \azero(980) or \azero(1450) and \fzero\ is 
\fzero(980).  Measurements of $B$ decays involving a scalar meson are
interesting since they provide information on such $B$ decays and the
nature of scalar mesons.  Several $B$ decays involving scalar mesons have 
been observed, either with an 
\fzero(980) \cite{f01,f02} or \Kstar(1430) \cite{kst1430} in the final state.  

\begin{figure}[!htb]
\vspace{0.5cm}
 \includegraphics[angle=0,width=\linewidth]{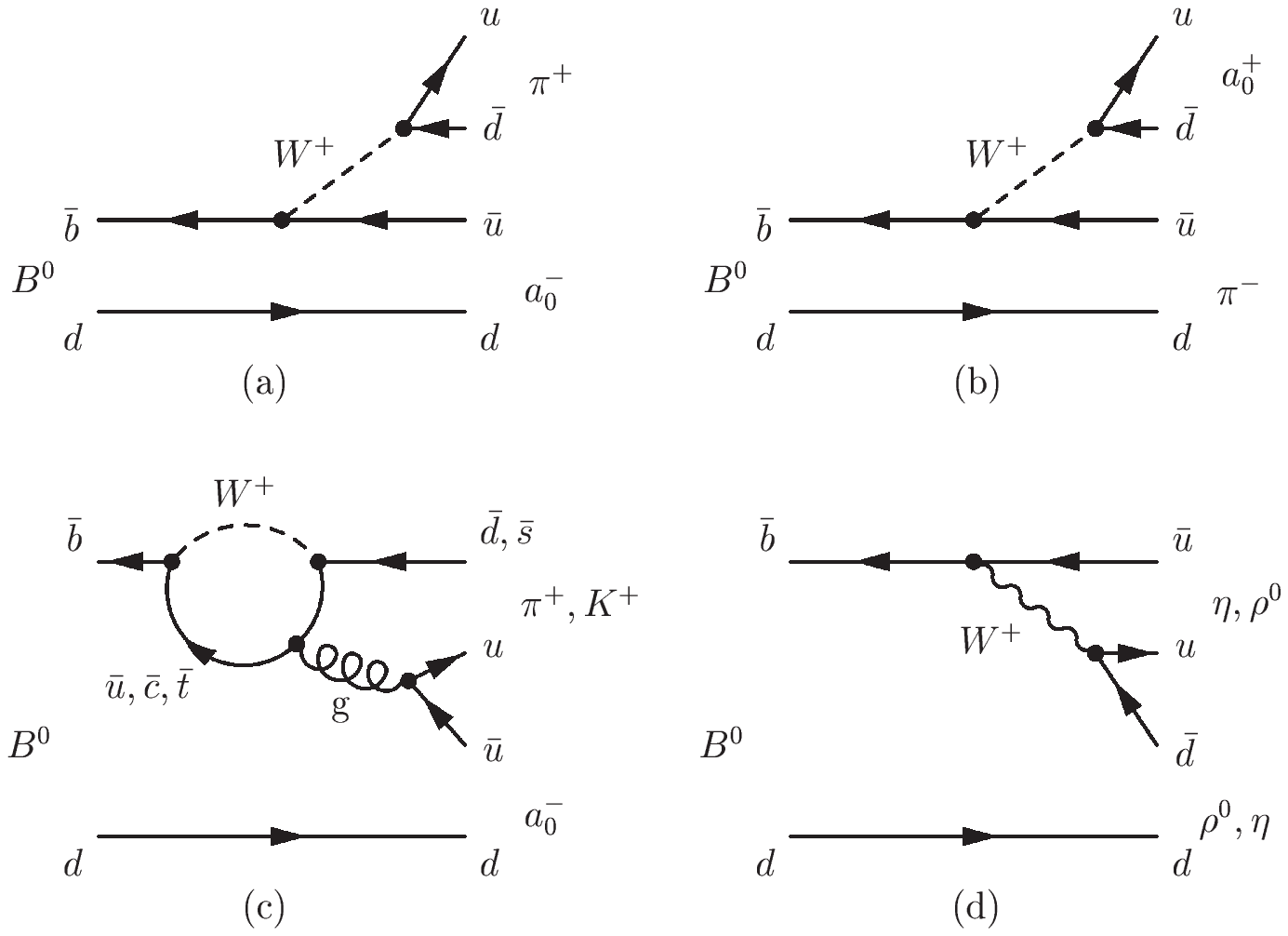}
 \caption{\label{fig:feyn}
Feynman diagrams for decays involving charged \azero\ mesons: (a) dominant and
(b) $G$-parity-suppressed tree diagrams for $\Bz\to\azero^\mp\pi^\pm$, 
(c) penguin diagram for $\Bz\to\azero^\mp\pi^\pm$ and \azeromKp\ (a tree
diagram similar to (a) also contributes to \azeromKp), (d) 
tree diagrams for \etarhoz\ (\etafz\ is similar with the \rhoz\
being replaced by a \fzero).}
\end{figure}

Specific predictions can be made for the decays $B\to\azero\pi$ if 
factorization is assumed and if the decay involves only tree or penguin (loop) 
processes.  The dominant amplitude is shown in
Fig.~\ref{fig:feyn}(a).  The companion tree amplitude, shown in
Fig.~\ref{fig:feyn}(b), is expected to be greatly suppressed, since,
neglecting light-quark mass splittings, the
virtual $W$ cannot produce an \azero\ meson \cite{vasia}.  This is a
firm prediction of the Standard Model because the weak current has a
$G$-parity even vector part and a $G$-parity odd axial-vector part. The
latter can produce an axial-vector or pseudoscalar particle
while the former produces a vector particle, but neither can produce a 
$G$-parity odd scalar meson ({\it e.g.} \azero).  Thus the
decay $B\to\azero\pi^\pm$ is expected to be ``self-tagging" (the charge of
the pion identifies the $B$ flavor).  
Penguin processes such as shown in Fig.~\ref{fig:feyn}(c) are allowed, but 
are suppressed relative to the tree processes.  
The decays with a kaon in the final state 
should be dominated by the penguin processes (Fig.~\ref{fig:feyn}(c)), though 
there is a cancellation between two terms in the penguin amplitudes for these 
decays \cite{chernyak}.  

The theoretical expectations for these decays
with an \azero(980) meson \cite{chernyak,minkochs,cheng,shen} are larger than 
previous experimental limits \cite{BABARa0}.  The decays with a \rhoz\
meson are expected to have branching fractions $\lsim1\times10^{-7}$
\cite{etarho} since they are dominated by color-suppressed tree amplitudes 
(Fig.~\ref{fig:feyn}(d)).  There are no predictions for the decay \etafz,
but it should have a small branching fraction for the same reason.

The nature of the \azero\ meson is not well understood.  It is
thought to be a \qqbar\ state with a possible admixture of a $K\Kbar$
bound-state component due to the proximity to the $K\Kbar$ threshold
\cite{PDG2006,baru}.  The \azero\ mass is known to be about 985 MeV with a 
width of $71\pm7$ \mev \cite{teige} for the dominant 
$\azero\to\eta\pi$ decay mode \cite{PDG2006}.  The \azero(1450) has a measured
width of 265 \mev \cite{PDG2006}.  Since the branching fraction for 
$\azero\to\eta\pi$ is not well known, we report the product branching 
fraction $\calB(\Bz\to\azero^- X^+)\times\calB(\azero\to\eta\pi)$,
where $X$ indicates $K$ or $\pi$.  The properties of the \fzero(980)
meson are well measured for the \pipi\ channel that is used in this
analysis \cite{aitala}.

Only limits for the following decays have been reported previously:
$\Bz\to a_0(980)^-\pi^+$ and $\Bz\to a_0(980)^-K^+$ \cite{BABARa0} and 
\etarhoz\ \cite{BABARrho,Bellerho,CLEOrho}.  There have been no previous 
searches for \etafz\ or the decays with an \azero(1450).  
The results presented here are based on data collected
with the \babar\ detector~\cite{BABARNIM}
at the PEP-II asymmetric-energy $e^+e^-$ collider
located at the Stanford Linear Accelerator Center.  An integrated
luminosity of approximately 347~fb$^{-1}$, corresponding to 
383\timesix\ \BB\ pairs, was recorded at the $\Upsilon (4S)$
resonance (center-of-mass energy $\sqrt{s}=10.58\ \gev$).

The track parameters of charged particles are measured by
a combination of a silicon vertex tracker, with five
layers of double-sided silicon sensors, and a
40-layer central drift chamber, both operating in the 1.5T magnetic
field of a superconducting solenoid. We identify photons and electrons 
using a CsI(Tl) electromagnetic calorimeter (EMC).  Further charged particle 
identification (PID) is provided by measurements of the average energy
loss (\dedx) in the tracking devices and by an internally-reflecting,
ring-imaging Cherenkov detector (DIRC) covering the central region.

We select $\azero$ candidates from the decay channel $\azero\to\eta\pi$ with 
the decays \etatogg\ (\etagg) and \etatoppp\ (\etappp).  All charged
tracks are required to originate from a common vertex.  For the decays
\etarhoz\ and \etafz, we use the same $\eta$ decay modes and $\rhoz\to\pipi$ 
or $\fzero\to\pipi$.  We apply the following 
requirements on the invariant masses (in \mev) relevant here:
$500< m_{\gaga}<585$ for \etagg, $535<m_{\pi\pi\pi}< 560$ for
\etappp, $120 < m_{\gaga} < 150$ for \piz, and $510<m_{\pi\pi}< 1060$
for $\rhoz/\fzero$.  For the \azero(980) analysis we use a mass range
$775 < m_{\eta\pi} < 1175$, while for \azero(1450) we require
$775 < m_{\eta\pi} < 1750$ (the latter upper limit removes unwanted background
from $D$ decays).  These requirements, except for \piz, are quite loose 
compared with typical resolutions in order to achieve high 
efficiency and retain sufficient sidebands to characterize the background for 
subsequent fitting.

We make several PID requirements to ensure the identity of the pions and kaons.
Secondary tracks from $\eta,~\rhoz$, or \fzero\ decays must have measured DIRC, 
\dedx, and EMC outputs consistent with pions.  For the \azerompip\ (\azeromKp) 
decays, we require an associated DIRC Cherenkov angle
between $-2$ and $+5$ ($-5$ and $+2$) standard deviations ($\sigma$) from the
expected value for a pion (kaon); the requirement is more restrictive on
the side where there is kaon (pion) background.  The distributions are corrected
as a function of momenta and angles so that the distributions are normalized
Gaussians.

A $B$-meson candidate is characterized kinematically by the
energy-substituted mass $\mes=(\frac{1}{4}s-\pvec_B^2)^\half$
and energy difference $\DE = E_B-\half\sqrt{s}$, where
$(E_B,\pvec_B)$ is the $B$-meson 4-momentum vector
and all values are calculated in the \UfourS\ frame.
Signal events peak at zero for \DE , and at the $B$ nominal mass for \mes .
The \DE\ (\mes) resolution is about 30 MeV ($3.0\ \mev$). 
We require $|\DE|\le0.2$ GeV and $5.25\le\mes\le5.29\ \gev$.

Backgrounds arise primarily from random combinations in continuum 
$\epem\ra\qqbar$ ($q=u,d,s,c$) events. We reduce these by using the angle
\thetaT\ between the thrust axis of the \Bz\ candidate in the \UfourS\
frame and that of the 
rest of the charged tracks and neutral clusters in the event.
The distribution of $|\costhr|$ is
sharply peaked near $1.0$ for combinations drawn from jet-like \qqbar\
pairs, and nearly uniform for $B$-meson decays.  We require $|\costhr|<0.7$
(0.8) for the \etagg\ (\etappp) channels.  We also use, in the fit 
described below, a Fisher discriminant \xf\ that combines the angles of
the candidate direction
with respect to the beam axis of the $B$ momentum and $B$ thrust axis 
(in the \UfourS\ frame), and moments describing 
the energy flow about the $B$ thrust axis \cite{PRD}.

We use additional event-selection criteria to further reduce backgrounds from 
charmless $B$ decays.  For the \etatogg\ modes we require
$|\cos{\theta^{\eta}_{\rm dec}}| \le 0.86 $, where $\theta^{\eta}_{\rm dec}$ is 
the angle of the photons in the $\eta$ rest frame with respect to the direction
of the particle recoiling against the $\eta$.  We also require
$|\cos{\theta^{a_0}_{\rm dec}}| \le 0.8$, where $\theta^{a_0}_{\rm dec}$ is 
defined similarly to $\theta^{\eta}_{\rm dec}$.  For \etarhoz\ decays, 
we define \hel\ to be the magnitude of the cosine of the angle
between the pion from the $\rho$ and the \Bz\ momentum in the $\rho$ rest frame,
and require $\hel<0.75$ to remove background from $\Bz\to D^+(\eta\pip)\pim$.
These additional requirements reduce the backgrounds by a
factor of 2--4, depending on the decay mode.  We use Monte Carlo (MC) 
simulation~\cite{geant} for an estimate of the residual \BB\ background 
and to identify the few 
(mostly charmless) decays that survive the candidate selection and have 
characteristics similar to the signal (20---270 events, depending on mode).
We include a component in the fit to account for them.

We obtain yields and branching fractions from extended unbinned 
maximum-likelihood fits, with input observables \DE, \mes, \mres\ for
the \azero(1450) fits, these observables plus \xf\ for the \azero(980)fits, 
and these plus \hel\ for the \fetarhoz/\fetafz\ fits.  The observable 
\mres\ denotes the $\eta\pi$ mass for the \azero\ analyses and the 
$\eta$ and \pipi\ masses for the \fetarhoz/\fetafz\ fits.  We employ
separate fits to determine the \azero(980) and \azero(1450) yields since
this results in $\sim$20\% better sensitivity for \azero(980).  The
\azero(1450) fit has a component for \azero(980) with the yield fixed to the 
value found in the \azero(980) fit, corrected for the small efficiency
difference.  No \azero(1450) component is used for the \azero(980) fit
since none of the subsamples have evidence of \azero(1450) signals.

For each event $i$ and hypothesis $j$ (signal, continuum background, 
\BB\ background), we define a product of probability density functions (PDF)
\begin{equation}
\calP^i_j=\calP_j (\mes^i) \calP_j (\DE^i) \calP_j (\mres^i) 
\left[\calP_j(\xf^i)\right] \left[\calP_j(\hel^i)\right] .
\end{equation}
The bracketed variables \xf\ and \hel\ are not used in all fits.
For all decays except for those involving \azero(1450), the absence of 
significant correlations among observables in the background is 
confirmed with the background-dominated data samples entering the fits.  
For the \azero(1450) decays, we find a substantial correlation in data
between \xf\ and the $\eta\pi$ invariant mass, due to the large $\eta\pi$
mass range.  We therefore require $\xf<0$ for these modes, which
keeps 90\% of the signal, and exclude \xf\ from the fit.  For the signal
component, we correct for effects due to the neglect of small correlations 
(more details are provided in the systematics discussion below).  The
\BB\ background yield is free in the \fetarhoz/\fetafz\ fits and is found 
to be in agreement with expectations from MC simulations.  For the decays
involving \azero\ mesons, we fix the \BB\ yield to the value predicted by MC
and include the uncertainty in the systematic errors (see below).

The likelihood function is
\begin{equation}
{\cal L} = \exp{(-\sum_j Y_j)}
\prod_i^{N}\left[\sum_j Y_j \calP^i_j\right]\,,
\end{equation}
where $Y_j$ is the yield of events of hypothesis $j$ that 
we find by maximizing \calL, and $N$ is the number of events in the sample. 

We determine the PDF parameters from simulation for the
signal and \BB\ background components.
We parameterize each of the functions $\calP_{\rm sig}(\mes),\ 
{\cal  P}_{\rm sig}(\DE),~\calP_j(\xf),~\rm and~\calP_{\rm sig}(\mres)$
with either the sum of two Gaussian functions, a Breit-Wigner shape, or an 
asymmetric Gaussian function, as required to describe the distribution.  
$\calP_{\rm sig}(\hel)$ for \etarhoz\ is described by a second order polynomial.
The shape of the real-meson component of $\calP_j(\mres)$ in the 
combinatorial background is described with the same parameters as for 
signal.  The distributions of \mres, \DE, and \hel\ for \BB\ and 
combinatorial background are represented by second order Chebyshev 
polynomials and/or the sum of two Gaussian functions.  
The \qqbar\ combinatorial background in \mes\
is described by the function $f(x)=x\sqrt{1-x^2}\exp{\left[-\xi(1-x^2)\right]}$,
with $x\equiv2\mes/\sqrt{s}$ and free parameter $\xi$; for peaking
\BB\ background, generally with the same or similar final state as signal,
we add a Gaussian function to the quantity $f(x)$.  

Large control samples 
of $B\to D\pi$ with a topology similar to the signal are used to verify 
the simulated resolutions in \DE\ and \mes.  Where the control data samples 
reveal small differences from MC, we shift or scale the resolution used in the 
likelihood fits.  Examples of many of these PDF shapes from a 
similar analysis are shown in Ref. \cite{PRD}.
Additionally, the signal parameters for the \azero(980) mass (983.5
\mev) and width (80 \mev)
are determined from an inclusive dataset that is much larger than the 
sample used for this analysis; they are consistent with
expectations from the natural-width values of Ref.~\cite{teige}.  The
values for \azero(1450) are taken from Ref.~\cite{PDG2006}.

In Table \ref{tab:results} we show for each decay mode the measured product 
branching fraction, together with the quantities entering into its 
determination.  In order to account for the uncertainties in the background
PDF parameterization, we include as free parameters in the fit, in addition to 
the signal and background yields, the principal 
parameters describing the background PDFs.  These include slopes for the
polynomial shape for the \DE\ and \mres\ distributions, the
parameter $\xi$ used for \mes, and three parameters
describing the asymmetric Gaussian function for \xf.
For calculation of branching fractions, we assume that the decay rates 
of the \UfourS\ to \BpBm\ and \BzBzb\ are equal \cite{prodratio}.
We combine branching fraction results from the two $\eta$ decay
channels by adding the values of $-2\ln{\cal L}$, adjusted for a
fit bias (see below) and taking proper account of the correlated and 
uncorrelated systematic errors.  We quote 90\% confidence level (C.L.)
upper limits, taken to be the branching fraction below which lies 90\% of
the total of the likelihood integral in the positive branching fraction
region.

\begin{figure}[!htb]
 \includegraphics[angle=0,width=\linewidth]{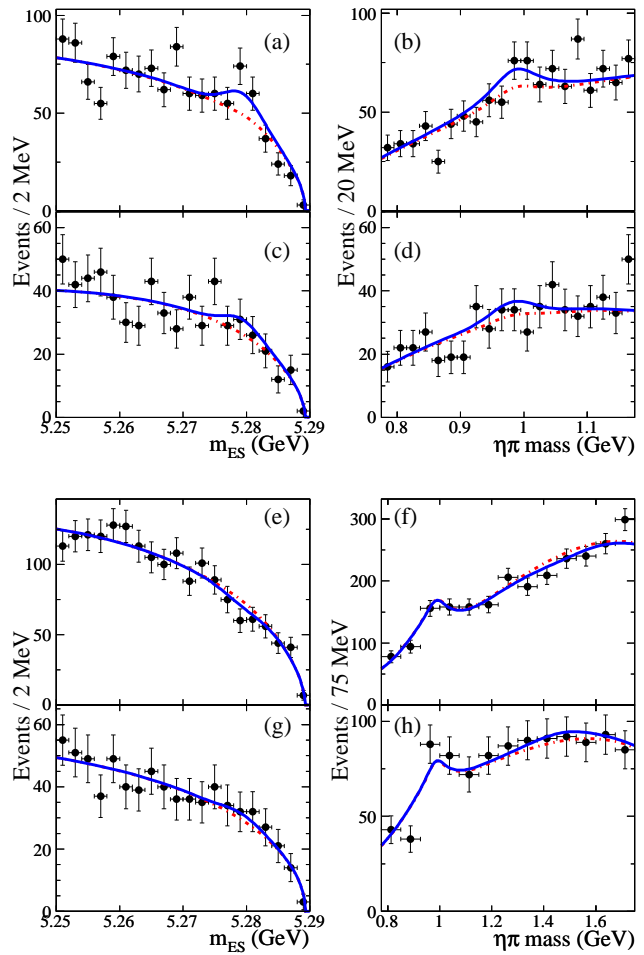}
 \caption{\label{fig:projazero}
Signal-enhanced projections of the \Bz-candidate \mes\ and $\eta\pi$ mass for (a, b) 
\fazerompip, (c, d) \fazeromKp, (e, f) \fazeropmpip, and (g, h) \fazeropmKp.  
Points with errors represent data, solid curves the full fit functions (both
signal modes combined), and  dot-dashed curves the background functions (the 
peaking \BB\ background component is small). 
For the \azero(1450) plots, the \azero(980) signal is
included in the background curves.  These plots are made with a minimum
requirement on the likelihood that has an efficiency for signal of 60--70\%.
}
\end{figure}

\begin{figure}[!htb]
 \includegraphics[angle=0,width=\linewidth]{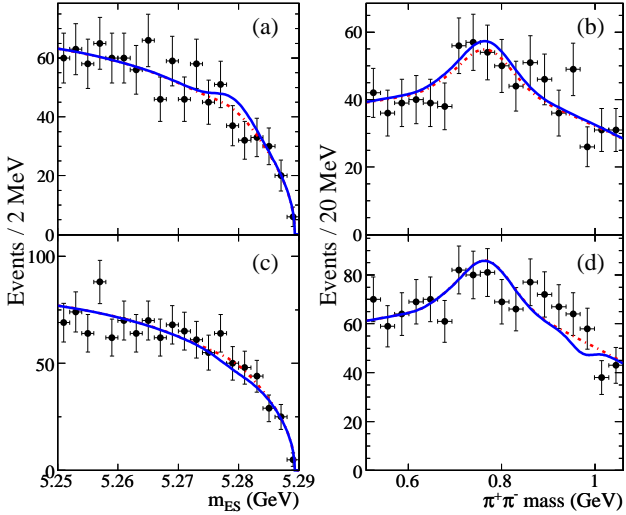}
 \caption{\label{fig:projrhofz}
Signal-enhanced projections of the \Bz-candidate \mes\ and \pipi\ mass for (a, b) \fetarhoz\ 
and (c, d) \fetafz. Points with errors represent data, solid curves the full fit
functions (both signal modes combined), and dot-dashed curves the background 
functions (the peaking \BB\ background component is small).  These plots
are made with a minimum requirement on the likelihood that has an efficiency 
for signal of 70--90\%.
}
\end{figure}

In Fig.~\ref{fig:projazero} we show projections onto 
\mes\ and $\eta\pi$ mass of subsamples enriched by a mode-dependent threshold 
requirement on the ratio of signal to total likelihood (computed without
the variable plotted).  We show analogous projections of \mes\ and the
\pipi\ invariant mass in Fig.~\ref{fig:projrhofz}.

\begin{table*}[!btp]
\caption{
Signal yield with statistical error, fit bias, detection efficiency $\epsilon$,
relevant daughter branching fraction product $\prod\calB_i$, significance 
(including additive systematic uncertainties), measured branching fraction 
\calB, and the 90\% C.L. upper limit on this branching fraction.  For the
\azero\ and \fzero\ modes, \calB\ includes the daughter 
branchings fractions for $\azero\to\eta\pi$ or $\fzero\to\pipi$.  }
\label{tab:results}
\begin{tabular}{lcrccccc}
\dbline
Mode	      & Fit yield& Bias$~~~~$& $~~~~\epsilon~~~~$&$~~\prod\calB_i~~$& ~Signif.~ & \calB & ~~\calB\ U.L.~~ \\
              &(events)&$~$(events)$~$& (\%) & (\%) & ($\sigma$) & ($10^{-6}$) & ($10^{-6}$) \\
\tbline
\bma{a_0(980)^-\pi^+} & & & & &\bma{\sazerompip}&\bma{\razerompip}&\bma{<\ulazerompip}\\
~~$a_0(980)^-_{\gaga}\pip$  & $87\pm23$   &$16~~~~~$& 15.3&39.4&3.3&$\msp3.1^{+1.1}_{-1.0}$\\
~~$a_0(980)^-_{3\pi}\pip$ &$4^{+12}_{-10}$&$ 1~~~~~$& 11.8&22.6&0.1&$\msp0.1^{+1.3}_{-1.1}$\\
\bma{a_0(980)^-\Kp}  & & & & &\bma{\sazeromKp} &\bma{\razeromKp} &\bma{<\ulazeromKp}\\
~~$a_0(980)^-_{\gaga}\Kp$ &$28^{+15}_{-13}$&$14~~~~~$& 14.0&39.4&2.0&$\msp1.1^{+0.7}_{-0.6}$\\
~~$a_0(980)^-_{3\pi}\Kp$ &$9^{+12}_{-9}$  &$11~~~~~$& 11.0&22.6&0.7&$\msp0.7^{+1.2}_{-1.0}$\\
\bma{\fazeropmpip}& & & & & ---              &\bma{\razeropmpip}&\bma{<\ulazeropmpip}\\
~~\fazeropmpipgg  &$-47\pm56$&$26~~~~~$& 13.8&39.4&--- &$-3.5^{+2.7}_{-2.6}$\\
~~\fazeropmpipppp &$-24\pm32$&$5~~~~~$ &  9.8&22.6&--- &$-3.5^{+3.8}_{-3.6}$\\
\bma{\fazeropmKp}& & & & &\bma{\sazeropmKp} &\bma{\razeropmKp} &\bma{<\ulazeropmKp}\\
~~\fazeropmKpgg  &$22\pm36$ &$12~~~~~$& 13.3&39.4&0.3&$\msp0.5^{+1.8}_{-1.7}$\\
~~\fazeropmKpppp &$13\pm24$ &$0~~~~~$ &  9.7&22.6&0.6&$\msp1.6^{+2.9}_{-2.7}$\\
\bma{\fetarhoz}& & & & &\bma{\setarhoz} &\bma{\retarhoz} &\bma{<\uletarhoz}\\
~~\fetaggrhoz  &$15^{+13}_{-11}$&$7~~~~~$& 10.7&39.4&0.7&$\msp0.5^{+0.8}_{-0.7}$\\
~~\fetappprhoz &$4^{+12}_{-10}$ &$5~~~~~$&  8.4&22.6&--- &$-1.4^{+1.6}_{-1.4}$\\
\bma{\fetafz}  & & & & & ---          &\bma{\retafz} &\bma{<\uletafz}\\
~~\fetaggfz    &$-11^{+10}_{-8}$&$1~~~~~$& 18.8&39.4&--- &$-0.4\pm0.3$\\
~~\fetapppfz   &$-4\pm11$       &$-4~~~~~$&14.9&22.6&0.1&$\msp0.0\pm0.5$\\
\dbline
\end{tabular}
\vspace{-5mm}
\end{table*}

The significance is taken as the square root of the difference between 
the value of $-2\ln{\cal L}$ (with additive systematic 
uncertainties included) for zero signal and the value at the minimum,
with other parameters free in both cases.

Most of the yield uncertainties arising from lack of knowledge of the PDFs have
been included in the statistical error since most background parameters are 
free in the fit.  Varying the signal PDF parameters within their estimated 
uncertainties, we determine the uncertainties in the signal yields to be 0--9 
events, depending on the final state.  This uncertainty is substantial only 
for the modes with one of the \azero\ resonances, where it is dominated by 
uncertainties in the parameterization of the \azero\ signal shape.
The neglect of correlations among observables in the fit can cause a systematic 
bias; the correction for this bias (between $-$4 and $+26$ events) and assignment
of the resulting systematic uncertainty (0.1--13 events) is determined from 
simulated samples with varying background populations.  For the \azero\
modes where the \BB\ background yield is fixed, we estimate the
uncertainty from modeling the \BB\ backgrounds by varying the expected
\BB\ yield by 100\% (0.1--12 events).

The above uncertainties are additive in nature and affect the
significance of the results.  Multiplicative uncertainties include our 
knowledge of the efficiency and other quantities entering the branching
fraction calculation.  Selection efficiency uncertainties are 1--2\% for
\costhr\ and 0.5--0.8\% due to the limited size of the MC samples.
Uncertainties in the reconstruction efficiency found from auxiliary studies 
on inclusive control samples~\cite{PRD}, include 0.5\%$\cdot N_t$ and 
1.5\%$\cdot N_\gamma$, where $N_t$ and $N_\gamma$ are the number of signal 
tracks and photons, respectively. The uncertainty on the total number of
\BB\ events is 1.1\%.  Published data
\cite{PDG2006}\ provide the uncertainties in the $B$-daughter product branching 
fractions (1--2\%).  

In conclusion, we do not find significant signals for the $B$-meson decays 
presented here.  The measured branching fractions and 90\% C.L. 
upper limits are given in Table \ref{tab:results}.  The limits for the
\azero(980) channels are smaller than expectations
\cite{chernyak,minkochs,cheng,shen}.  This has been cited as evidence
that the \azero(980) meson is a four-quark state, not the lowest-lying  
member of the \qqbar\ scalar multiplet \cite{cheng}.  The limits for the 
\azero(1450) channels and \etarhoz\ are consistent with theoretical 
expectations \cite{cheng,etarho}.  There are no previous measurements or
theoretical predictions for the \etafz\ decay.

We are grateful for the excellent luminosity and machine conditions
provided by our \pep2\ colleagues, 
and for the substantial dedicated effort from
the computing organizations that support \babar.
The collaborating institutions wish to thank 
SLAC for its support and kind hospitality. 
This work is supported by
DOE
and NSF (USA),
NSERC (Canada),
CEA and
CNRS-IN2P3
(France),
BMBF and DFG
(Germany),
INFN (Italy),
FOM (The Netherlands),
NFR (Norway),
MIST (Russia),
MEC (Spain), and
PPARC (United Kingdom). 
Individuals have received support from the
Marie Curie EIF (European Union) and
the A.~P.~Sloan Foundation.

\end{document}